# Nonlinear Observer Design and Synchronization Analysis for Classical Models of Neural Oscillators

Ranjeetha Bharath and Jean-Jacques Slotine
Massachusetts Institute of Technology

ABSTRACT

This work explores four nonlinear classical models of neural oscillators, the Hodgkin-Huxley model, the Fitzhugh-Nagumo model, the Morris-Lecar model, and the Hindmarsh-Rose model. Nonlinear contraction theory is used to develop observers and perform synchronization analysis on these systems. Neural oscillation and signaling models are based on the biological function of the neuron, with behavior mediated through the channeling of ions across the cell membrane. The variable assumed to be measured is the membrane potential, which may be obtained empirically through the use of a neuronal force-clamp system, or may be transmitted through the axon to other neurons. All other variables are estimated by using partial state or full state observers. Basic observer rate convergence analysis is performed for the Fitzhugh Nagumo system, partial state observer design is performed for the Morris-Lecar system, and basic synchronization analysis is performed for both the Fitzhugh-Nagumo and the Hodgkin-Huxley systems.

# 1. INTRODUCTION

The goal of this work is to analyze four different biological models from a control systems perspective by designing nonlinear observers for them and studying their synchronization properties. The four models are the Hodgkin-Huxley model, the Fitzhugh-Nagumo model, the Morris-Lecar model, and the Hindmarsh-Rose model. These four models describe neuronal activity by modeling neurons in an oscillatory fashion. Neuronal networks involve "networks of interactions between cells" and consist of "the network of synaptic connections between neurons."[1] The time scale for neuronal activity is milliseconds and the length scale is cellular.[1] Analyzing the mathematical behavior of neural networks can lead to interesting insights about cellular behavior and interaction.

The first section describes a set of mathematical notions underlying the theory presented in this work, such as contraction analysis. The next section discusses the biological motivation and mathematical analysis of the Hodgkin-Huxley model developed as a four-state description of signaling. The third section describes the Fitzhugh-Nagumo model, which is a two-state simplification of the Hodgkin-Huxley model. The fourth section describes the Morris-Lecar model, which is a three-state description. Finally, the fifth section describes the Hindmarsh-Rose model, which is another three state model description. In all cases, for observer design, the membrane potential or voltage is taken to be the measured variable. These sections are followed by a conclusion and references.

# 2. MATHEMATICAL NOTIONS: CONTRACTION[12]

The concept of contraction can be made rigorous by studying the system mathematically. The deterministic system, which can be multivariable and nonlinear in any and all of the state variables, is given as:

$$\dot{x} = f(x,t), with$$
$$\delta\dot{x} = \frac{\partial f}{\partial x}(x,t)\delta x$$

It is valuable to define what is meant, in matrix algebra, by uniformly negative definite before proceeding with the system analysis. To say that $\frac{\partial f}{\partial x}(x,t)$ is uniformly negative definite is to say that

$$\exists \beta > 0, \forall x, \forall t \geq 0, \frac{1}{2}\left(\frac{\partial f}{\partial x} + \frac{\partial f^T}{\partial x}\right) \leq -\beta I < 0$$

However, a simple application of this idea can be used in the following manner instead[14]: a virtual system constructed as a copy of the real system which accounts for the coupling "is contracting if the maximum eigenvalue of the symmetric part of F, where F is the generalized Jacobian, is uniformly negative." This is the notion used in the remainder of this work in order to study the synchronization behavior of coupled neurons.

# 3. HODGKIN-HUXLEY[5-8]

3.1 BIOLOGICAL MOTIVATION
By studying the electrical behavior of the giant axon in squid, A. L. Hodgkin and A. F. Huxley were able to create a mathematical model describing neuronal activity in 1955. Naturally, this model is a simplification of the significantly more complicated real neural potential dynamics. The three current components, which together make up the basis for ion transfer across the membrane, are $I_{Na}$, $I_K$, and $I_l$, referring to sodium, potassium, and other ions, respectively. The sodium and potassium conductances, $g_{Na}$ and $g_K$, are functions of time, but other factors such as ionic driving gradients (e.g. $E_{Na}$), are viewed as constant. It is proposed that the membrane potential controls permeability.

$$I_i = I_{Na} + I_K + I_l = g_{Na}(V - V_{Na}) + g_K(E - E_K) + g_l(E - E_l)$$

The paper finds equations to describe conductances, and then tries to provide a physical basis for the equations, since it claims that "there is little hope of calculating the time course of the sodium and potassium conductances from first principles."[4] The variable n is used as a dimensionless variable representing the proportion of particles inside versus outside the membrane. It varies between 0 and 1. In a like manner, m and h are variables used to represent proportions and range from 0 to 1 as well. The Hodgkin-Huxley variable ranges are given in Table 1.

**Table 1**: Range for variables V, m, n, and h.

| Variable | V (mV) | m | n | h |
|---|---|---|---|---|
| Range | ~-110-110 | 0 to 1 | 0 to 1 | 0 to 1 |

3.2 MATHEMATICAL ANALYSIS
The Hodgkin-Huxley model is mathematically described by the following differential equations:

$$\dot{V} = \frac{I}{C_M} + n^4(V_k - V)\frac{\overline{g_k}}{C_M} + m^3h(V_{Na} - V)\frac{\overline{g_{Na}}}{C_M} + (V_{Na} - V)\frac{\overline{g_h}}{C_M}$$
$$\dot{m} = \alpha_m(V)(1 - m) - \beta_m(V)m$$
$$\dot{n} = \alpha_n(V)(1 - n) - \beta_n(V)n$$
$$\dot{h} = \alpha_h(V)(1 - h) - \beta_h(V)h$$
$$\alpha_m = \frac{0.1(V + 25)}{e^{\frac{V+25}{10}} - 1}, \beta_m = 4e^{V/18}$$
$$\alpha_n = \frac{0.01(V + 10)}{e^{\frac{V+10}{10}} - 1}, \beta_n = 0.125e^{V/80}$$
$$\alpha_h = 0.07e^{V/20}, \beta_h = \frac{1}{e^{\frac{V+30}{30}} + 1}$$

Using the model description above and differential calculus for partial differential equations, it thus follows that the Jacobian of the entire system is:

$$\begin{bmatrix}
-n^4\frac{g_k}{C_M}-m^3h\frac{g_{Na}}{C_M}-\frac{1}{C_M}\frac{g_h}{C_M} & 4n^3(V_k-V)\frac{g_k}{C_M} & 3m^2(V_{Na}-V)\frac{g_{Na}}{C_M} & m^3(V_{Na}-V)\frac{g_{Na}}{C_M} \\
\frac{\left(0.1\left(e^{\frac{V+25}{10}}-1\right)-0.01(e^{\frac{V+25}{10}}-1)e^{\frac{V+25}{10}}\right)(1-m)}{\left(e^{\frac{V+25}{10}}-1\right)^2} - \frac{2}{9}e^{V/18}m & \frac{-0.1(V+25)}{e^{\frac{V+25}{10}}-1}-4e^{V/18} & 0 & 0 \\
\frac{\left(0.01\left(e^{\frac{V+10}{10}}-1\right)-0.001(e^{\frac{V+10}{10}}-1)e^{\frac{V+10}{10}}\right)(1-n)}{\left(e^{\frac{V+10}{10}}-1\right)^2} - \frac{0.125}{80}e^{V/80}n & 0 & \frac{-0.01(V+10)}{e^{\frac{V+10}{10}}-1}-0.125e^{V/80} & 0 \\
\left(\frac{0.07}{20}e^{\frac{v}{20}}\right)(1-h)+\frac{e^{\frac{v+30}{10}}h}{\left(e^{\frac{v+30}{10}}+1\right)^2} & 0 & 0 & -0.07e^{\frac{V}{20}}+\frac{1}{e^{\frac{V+30}{10}}+1}
\end{bmatrix}$$

### 3.3 OBSERVER DESIGN

In reality, not all these variables might be measurable with physical system. Assuming that only the membrane potential, V, can be measured, it is valuable to construct an observer in order to fill in for the state variables that cannot be measured but are important to describe the system.

Replace the system with a partial state observer, copying the m, n, and h dynamics. Then, assume that the membrane potential is the only measured variable. This line of analysis yields the following results:

$$\dot{\hat{m}} = \frac{0.1(V+25)}{e^{\frac{V+25}{10}}-1}(1-\hat{m})-4e^{V/18}\hat{m}$$

$$\dot{\hat{n}} = \frac{0.01(V+10)}{e^{\frac{V+10}{10}}-1}(1-\hat{n})-0.125e^{V/80}\hat{n}$$

$$\dot{\hat{h}} = 0.07e^{V/20}(1-\hat{h})-\frac{1}{e^{\frac{V+30}{30}}+1}\hat{h}$$

Taking the derivatives of the observer dynamics with respect to the estimated variables yields:

$$\frac{d\dot{\hat{m}}}{d\hat{m}} = -\frac{0.1(V+25)}{e^{\frac{V+25}{10}}-1}-4e^{V/18}$$

$$\frac{d\dot{\hat{n}}}{d\hat{n}} = -\frac{0.01(V+10)}{e^{\frac{V+10}{10}}-1}-0.125e^{V/80}$$

$$\frac{d\dot{\hat{h}}}{d\hat{h}} = -0.07e^{\frac{V}{20}}-\frac{1}{e^{\frac{V+30}{30}}+1}$$

The diagonal Jacobian of the observer is thus:

$$\begin{bmatrix} -\dfrac{0.1(V+25)}{e^{\frac{V+25}{10}}-1} - 4e^{V/18} & 0 & 0 \\ 0 & -\dfrac{0.01(V+10)}{e^{\frac{V+10}{10}}-1} - 0.125e^{V/80} & 0 \\ 0 & 0 & -0.07e^{\frac{V}{20}} - \dfrac{1}{e^{\frac{V+30}{30}}+1} \end{bmatrix}$$

It is not difficult to show the range of the diagonal terms for the domain of V. Plotting the (1,1) and (2,2) terms as functions of V reveal their ranges. The (1,1) and (2,2) terms are negative for all values of V, since the negative exponential value offsets the cases when V is less than -10 or -25. The (3,3) entry is obviously negative for any value of V, since the range of the exponential function ($e^x$) is strictly positive.

3.4 COUPLING AND SYNCHRONIZATION ANALYSIS

The next step is to study the possible coupling of multiple Hodgkin-Huxley neurons.

$$\dot{V} = \frac{I}{C_M} + n^4(V_k - V)\frac{\overline{g_k}}{C_M} + m^3 h(V_{Na} - V)\frac{\overline{g_{Na}}}{C_M} + (V_{Na} - V)\frac{\overline{g_h}}{C_M} + \sum_{j \in N_i} k_{ji}(V_j - V_i)$$

$$\dot{m} = \alpha_m(V)(1 - m) - \beta_m(V)m$$
$$\dot{n} = \alpha_n(V)(1 - n) - \beta_n(V)n$$
$$\dot{h} = \alpha_h(V)(1 - h) - \beta_h(V)h$$
$$\alpha_m = \frac{0.1(V + 25)}{e^{\frac{V+25}{10}} - 1}, \beta_m = 4e^{V/18}$$
$$\alpha_n = \frac{0.01(V + 10)}{e^{\frac{V+10}{10}} - 1}, \beta_n = 0.125e^{V/80}$$
$$\alpha_h = 0.07e^{V/20}, \beta_h = \frac{1}{e^{\frac{V+30}{30}} + 1}$$

In the case of two neurons, there are eight state variables and equations, displayed as:

$$\dot{V_1} = \frac{I}{C_M} + n_1^4(V_k - V_1)\frac{\overline{g_k}}{C_M} + m_1^3 h_1(V_{Na} - V_1)\frac{\overline{g_{Na}}}{C_M} + (V_{Na} - V_1)\frac{\overline{g_h}}{C_M} + \sum_{j \in N_i} k_{21}(V_2 - V_1)$$

$$\dot{V_2} = \frac{I}{C_M} + n_2^4(V_k - V_2)\frac{\overline{g_k}}{C_M} + m_2^3 h_2(V_{Na} - V_2)\frac{\overline{g_{Na}}}{C_M} + (V_{Na} - V_2)\frac{\overline{g_h}}{C_M} + \sum_{j \in N_i} k_{21}(V_1 - V_2)$$

$$\dot{m_1} = \alpha_m(V_1)(1 - m_1) - \beta_m(V_1)m_1$$
$$\dot{n_1} = \alpha_n(V_1)(1 - n_1) - \beta_n(V_1)n_1$$
$$\dot{h_1} = \alpha_h(V_1)(1 - h_1) - \beta_h(V_1)h_1$$
$$\dot{m_2} = \alpha_m(V_2)(1 - m_2) - \beta_m(V_2)m_2$$
$$\dot{n_2} = \alpha_n(V_2)(1 - n_2) - \beta_n(V_2)n_2$$
$$\dot{h_2} = \alpha_h(V_2)(1 - h_2) - \beta_h(V_2)h_2$$

This leads to a very complicated Jacobian of size 8 by 8. However, this is a very similar Jacobian to the original 4 by 4 Jacobian of the uncoupled system. The matrix will still have similar diagonal and first column and row terms as the form of the equations has not changed, except for the addition of a $k_{21}(V_2 - V_1)$ and a $k_{21}(V_1 - V_2)$ term. Of course, if one were to make an observer for this expanded system, this coupling does not affect the reduced order observer, since these terms do not include m, n, and h.
The (1,1), (1,2), (2,1), and (2,2) terms are thus:

$$\begin{bmatrix} -n_1^4 \frac{\overline{g_k}}{C_M} - m_1^3 h_1 \frac{\overline{g_{Na}}}{C_M} - \frac{\overline{g_h}}{C_M} - k_{21} & k_{21} \\ k_{21} & -n_2^4 \frac{\overline{g_k}}{C_M} - m_2^3 h_2 \frac{\overline{g_{Na}}}{C_M} - \frac{\overline{g_h}}{C_M} - k_{21} \end{bmatrix}$$

This matrix just concerning $\frac{\partial V_1}{\partial V_1}, \frac{\partial V_1}{\partial V_2}, \frac{\partial V_2}{\partial V_1}$, and $\frac{\partial V_2}{\partial V_2}$ is symmetric. In order for both eigenvalues to be negative, the (1,1) and (2,2) diagonal terms must be larger in absolute value than the off diagonal terms, and yet more negative, through diagonal dominance. Since $-n_1^4 \frac{\overline{g_k}}{C_M} - m_1^3 h_1 \frac{\overline{g_{Na}}}{C_M} - \frac{\overline{g_h}}{C_M} - k_{21}$ contains $k_{21}$ and m, n, and h are always nonnegative ranging from 0 to 1, this will always be the case *except* for m, n, and h=0. The matrix will in this case reduce to:

$$\begin{bmatrix} -k_{21} & k_{21} \\ k_{21} & -k_{21} \end{bmatrix}$$

The eigenvalues will then be $-2k_{21}$ and 0, meaning that there is a loss in stability for this portion of the Jacobian. Similar analyses regarding the nature of submatrices of the rest of the 8x8 Jacobian can be used to better characterize the synchronization properties and see whether or not $V_1$ converges to $V_2$. In particular, it is suggested that by studying the four 4x4 matrices which compose the Jacobian, and seeing if they can be rearranged in combination (e.g. hierarchical combination[12]), further analysis can be performed.

# 4. FITZHUGH-NAGUMO MODEL[2]

4.1 BIOLOGICAL MOTIVATION
The Hodgkin-Huxley model proposed in 1952 can be approximated to two state variables instead of four, presented in the Fitzhugh-Nagumo model analysis. The Fitzhugh-Nagumo model attempts to characterize the threshold phenomenon in neuronal firing and signaling behavior. Note that a special case of this biological neural oscillator model is a simple van der Pol oscillator. Biologically, v represents the membrane potential just as it does in the parent Hodgkin-Huxley model. The variable w represents a recovery variable for the system behavior, and $I_{ext}$ refers to the external current applied.

4.2 MATHEMATICAL ANALYSIS
The model is described by the following set of two differential equations:
$$\dot{v}_i = c\left(v_i + w_i - \frac{1}{3}v_i^3 + I_{ext}\right) + \sum k_{ji}(v_j - v_i)$$
$$\dot{w}_i = -\frac{1}{c}(v_i - a + bw_i) \quad i = 1, \dots, n$$
It is easy to do a quick contraction analysis on this system after building an observer and using a transformation matrix to form a generalized Jacobian.

$$J = \begin{pmatrix} c(1-v^2) & c \\ -1/c & -b/c \end{pmatrix}$$

4.3 OBSERVER DESIGN

Replace the system with a copy.

$$J = \begin{pmatrix} c(1-\hat{v}^2) & c \\ -1/c & -b/c \end{pmatrix}$$

By using the transformation matrix $\theta$ and the generalized Jacobian $\dot{\theta}\theta^{-1} + \theta\frac{\partial f}{\partial x}\theta^{-1}$ we arrive at a transformed Jacobian.

$$\theta = \begin{pmatrix} 1 & 0 \\ 0 & c \end{pmatrix} \quad \theta J \theta^{-1} = \begin{pmatrix} c(1-\hat{v}^2) & 1 \\ -1 & -b/c \end{pmatrix}$$

The symmetric part of the Generalized Jacobian is simply:

$$\theta J \theta^{-1} = \begin{pmatrix} c(1-\hat{v}^2) & 0 \\ 0 & -b/c \end{pmatrix}$$

Constructing an observer, assuming that v is measured yields:

$$\dot{\hat{v}} = c\left(\hat{v} + w_1 - \frac{1}{3}\hat{v}^3\right) - k(\hat{v} - v)$$

$$\dot{w} = -\frac{1}{c}(\hat{v} - a + b\hat{w})$$

Feeding in the observer gain k gives:

$$\theta J \theta^{-1} = \begin{pmatrix} c(1-\hat{v}^2) - k & 0 \\ 0 & -b/c \end{pmatrix}$$

For simplicity in mathematical analysis, take a reasonable choice for c, c=1. Then, it is clear that k must be greater than or equal to 1, since the maximum value for the (1,1) coordinate is 1, since $x^2$ is nonnegative for all real x. Because v is real, $1-v^2$ will always be less than or equal to 1. Also, -b/c is always negative since both b and c are positive constants. Since the system is contracting, $\hat{v}$ will converge to v.

4.4 OBSERVER RATE CONTROL APPROACH

It is desirable to control both rows of the system in order to affect the rate of convergence for the observer. The first option to study is feeding back the v error to the second row of the symmetric part of the generalized Jacobian, in the form $k_2(\hat{v} - v)$. This yields:

$$\theta J \theta^{-1} = \begin{pmatrix} c(1-\hat{v}^2) - k & 0 \\ -k_2 & -b/c \end{pmatrix}$$

Sylvester's criterion (this is not a diagonal matrix) yields that it is symmetric negative definite because the 1x1 determinant is negative for proper choice in k, and the full determinant is negative is:

$$det = \frac{b}{c}(-c(1-\hat{v}^2) + k = -b(1-\hat{v}^2) + \frac{k_2}{c}$$

Of course, choosing $k_2$ appropriately yields a negative 2x2 determinant. The next step is to consider how to choose $k_1$ and $k_2$ to place both the eigenvalues of the system and thus control the *rate of convergence* of the observer, not just the fact that it does indeed converge. Unfortunately, the two eigenvalues are still the same as in the case without this feedback, so it is not valuable—$k_2$ still does not play a role.

Another option is to choose other theta transformation matrices. This was tried for many cases but it made the generalized Jacobian very complicated and difficult to analyze, without yielding much physical gain. The generalized Jacobian loses its anti-symmetric properties for other choices of θ than what is presented above.

4.5 COUPLING AND SYNCHRONIZATION ANALYSIS

It is then valuable to analyze the generalized Jacobian and dynamics for a coupled system. The uncoupled Jacobian for both neurons is given as:

$$J = \begin{pmatrix} c(1 - v_1^2) & 0 & c & 0 \\ 0 & c(1 - v_2^2) & 0 & c \\ -1/c & 0 & -b/c & 0 \\ 0 & -1/c & 0 & -b/c \end{pmatrix}$$

By using the transformation matrix $\theta$ and the generalized Jacobian $\dot{\theta}\theta^{-1} + \theta \frac{\partial f}{\partial x}\theta^{-1}$ we arrive at a transformed Jacobian.

$$\theta = \begin{pmatrix} 1 & 0 & 0 & 0 \\ 0 & 1 & 0 & 0 \\ 0 & 0 & c & 0 \\ 0 & 0 & 0 & c \end{pmatrix} \quad \theta J \theta^{-1} = \begin{pmatrix} c(1 - v_1^2) & 0 & -1 & 0 \\ 0 & c(1 - v_2^2) & 0 & -1 \\ 1 & 0 & -b/c & 0 \\ 0 & 1 & 0 & -b/c \end{pmatrix}$$

This transformed Jacobian is anti-symmetric or skew-symmetric, and the diagonal terms can be analyzed for contraction purposes.

Now, by appropriately adding controller gains $k_1$ and $k_2$:

$$J_{Coupled} = \begin{pmatrix} c(1 - v_1^2) - k_1 & k & -1 & 0 \\ k & c(1 - v_2^2) - k_2 & 0 & -1 \\ 1 & 0 & -b/c & 0 \\ 0 & 1 & 0 & -b/c \end{pmatrix}$$

appropriately chosen to be positive and greater than 1, the coupled system is contracting. $(1-v^2)$ cannot be greater than 1 because v is a real number so $v^2$ is positive.

If the coupling is added in:

$$\hat{v}_1 = c\left(v_1 + w_1 - \frac{1}{3}v_1^3 + I_{ext}\right) + \sum k_{21}(v_2 - v_1)$$
$$\hat{v}_2 = c\left(v_2 + w_2 - \frac{1}{3}v_2^3 + I_{ext}\right) + \sum k_{12}(v_1 - v_2)$$
$$\dot{w}_1 = -\frac{1}{c}(v_1 - a + bw_1)$$
$$\dot{w}_2 = -\frac{1}{c}(v_1 - a + bw_2)$$

$$J_{Coupled} = \begin{pmatrix} c(1-v_1^2) - \sum k - k_1 & k & -1 & 0 \\ k & c(1-v_2^2) - \sum k - k_2 & 0 & -1 \\ 1 & 0 & -b/c & 0 \\ 0 & 1 & 0 & -b/c \end{pmatrix}$$

This concludes a preliminary analysis of the Fitzhugh-Nagumo neural oscillator model from a nonlinear control systems perspective.

# 5. MORRIS-LECAR MODEL[10]

### 5.1 BIOLOGICAL MOTIVATION
In 1981 at the National Institutes of Health in Bethesda, Maryand, Catherine Morris and Harold Lecar developed a model related to the Hodgkin-Huxley model, but using barnacle muscle fibers subjected to stimulation and a different dynamical description. Their model, called the Morris-Lecar model, allows for $I_K$ and and $I_{Ca}$ to dominate the potential dynamics. The model is conducted in three variables: V, which is the same voltage from the Hodgkin-Huxley model, m, which is the fraction of open $Ca^{2+}$ models, and n, which is the fraction of open $K^+$ channels. These variables are analogous to the m and n in the Hodgkin-Huxley model, which were ratios of these ions inside and outside the cell membrane. In all the experiments that this model is based off of (at least those presented in the paper first describing the model), the voltage ranges from -50mV to 50 mV. In addition, m and n are defined to (as ratios) range from 0 to 1. Various other physical parameters are defined in the paper.

Table 2 gives the values of particular constants used in the Morris-Lecar model which are pertinent to this preliminary control systems analysis.

**Table 2**: Values for constants in the Morris-Lecar model.

| Constant | Value |
|---|---|
| $V_1$ | 10 mV |
| $V_2$ | 15 mV |
| $V_3$ | -1 mV |
| $V_4$ | 14.5 mV |
| $g_k$ | 0 to 20 |
| $g_{Ca}$ | 0 to 20 |
| C | 20 |

Figure 1 depicts plots from the original Morris-Lecar paper showing how the variables $g_k$ and $g_{Ca}$ vary with one another for different values of I.

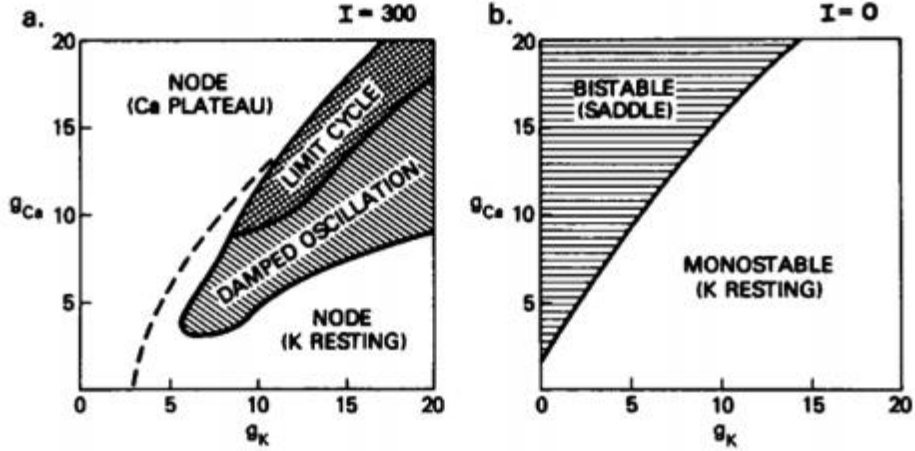

**Figure 1**. Variation of $g_{Ca}$ and $g_k$ with one another for varying values of $I^{10}$.

5.2 MATHEMATICAL ANALYSIS

The Morris-Lecar model can be described by the following set of three differential equations in three state variables:

$$C\dot{V} = I - g_L V_L - g_{Ca} m(V - V_{Ca}) - g_K n(V - V_K)$$

$$\dot{m} = \overline{\lambda_m} \cosh\left(\frac{V - V_1}{2V_2}\right)\left(\frac{1}{2}\left(1 + \tanh\left(\frac{V - V_1}{V_2}\right)\right) - m\right)$$

$$\dot{n} = \overline{\lambda_n} \cosh\left(\frac{V - V_3}{2V_4}\right)\left(\frac{1}{2}\left(1 + \tanh\left(\frac{V - V_3}{V_4}\right)\right) - n\right)$$

The Jacobian of this system is thus:

$$\begin{bmatrix} \frac{-g_{Ca}m}{c} - \frac{g_K n}{c} & \frac{-g_{Ca}V}{c} & \frac{-g_K V}{c} \\ \frac{\overline{\lambda_m}}{V_2}\left[\sinh\left(\frac{V-V_1}{2V_2}\right)\left(0.25\tanh\left(\frac{V-V_1}{V_2}\right) - \frac{m}{2} + \frac{1}{4}\right) + \frac{1}{2}\cosh\left(\frac{V-V_1}{2V_2}\right)\text{sech}^2\left(\frac{V-V_1}{V_2}\right)\right] & -\overline{\lambda_m}\cosh\left(\frac{V-m}{2V_2}\right) & 0 \\ \frac{\overline{\lambda_n}}{V_4}\left[\sinh\left(\frac{V-V_3}{2V_4}\right)\left(0.25\tanh\left(\frac{V-V_3}{V_4}\right) - \frac{n}{2} + \frac{1}{4}\right) + \frac{1}{2}\cosh\left(\frac{V-V_3}{2V_4}\right)\text{sech}^2\left(\frac{V-V_3}{V_4}\right)\right] & 0 & -\overline{\lambda_n}\cosh\left(\frac{V-n}{2V_4}\right) \end{bmatrix}$$

An analysis of individual Jacobian matrix entries yields interesting results for observer design. The (2,1) and (3,1) entries can be treated as:

$$\frac{\overline{\lambda_m}}{V_2}\left[\sinh\left(\frac{x}{2}\right)\left(0.25\tanh(x) - \frac{m}{2} + \frac{1}{4}\right) + \frac{1}{2}\cosh\left(\frac{x}{2}\right)\text{sech}^2(x)\right], \text{replacing}\left(\frac{V - V_1}{2V_2}\right) \text{with } x$$

It is useful to characterize the greatest value of these off-diagonal terms in order to study it from a matrix norm perspective of diagonal dominance to assess contraction properties of the system.

Experimentally, from values given in the original paper, the range of the x parameter given above can be determined. Using these given values, it is shown that the value of the x parameter varies approximately from $-2 \leq x \leq 2$. In addition, the m and n parameters, taken from the Hodgkin-Huxley model, are defined to vary from 0 to 1. Table 3 displays the values for the (2,1) and (3,1) terms for various x (column) and m (row) values.

**Table 3**: Values for the (2,1) and (3,1) terms, respectively, with maximum values given in bold.

| x \ m | 0 | 0.50 | 1 |
|---|---|---|---|
| -2 | 0.06 | 0.06 | 0.05 |
| -1 | 0.24 | 0.24 | 0.24 |
| 0 | **0.50** | **0.50** | **0.50** |
| 1 | 0.24 | 0.24 | 0.24 |
| 2 | 0.06 | 0.06 | 0.05 |

| x \ n | 0.00 | 0.50 | 1.00 |
|---|---|---|---|
| -2 | 0.06 | 0.06 | 0.05 |
| -1 | 0.24 | 0.24 | 0.24 |
| 0 | **0.50** | **0.50** | **0.50** |
| 1 | 0.24 | 0.24 | 0.24 |
| 2 | 0.06 | 0.06 | 0.05 |

The maximum values for the (2,1) and the (3,1) entries are given in bold. Recall that the diagonal dominance norm as it is used for contraction analysis purposes is given as

$$Diagonal\ Entry + \sum |Off\ Diagonal\ Terms| < 0$$

The (1,2) and (1,3) terms are bounded by constants, and can easily be off-set by a gain added to the diagonal term. Finally, the (2,3) and (3,2) terms are both zero.

Bounding the off-diagonal terms reduces the Jacobian analysis to simply:

$$\begin{bmatrix} \frac{-g_{Ca}m}{c} - \frac{g_K n}{c} & \leq 50 & \leq 50 \\ \leq 0.5 & -\overline{\lambda_m}\cosh(\frac{V-m}{2V_2}) & 0 \\ \leq 0.5 & 0 & -\overline{\lambda_n}\cosh(\frac{V-n}{2V_4}) \end{bmatrix}$$

Finally, looking at the diagonals, it is easy to see that the (1,1) term can be 0 if m and n are 0. The (2,2) and (3,3) entries can be studied by looking at the ranges of the cosh functions. The range of cosh is always positive.

$\overline{\lambda_m}$ and $\overline{\lambda_n}$ are both positive constants (1/15 and 1/10, respectively), and thus (2,2) and (3,3) are always negative. For the ranges of V for which this study and model is concerned with, (2,2) ranges from approximately -0.106 to -0.1045, and (3,3) ranges from the values -0.002267 to -0.00459. When choosing the gains, it is valuable to add a buffer for diagonal dominance. As is obvious, this diagonal dominance analysis for the full system is very complicated and also very conditional on the particular circumstances and current state values, which would need to be fed back to cancel out the varying terms. Thus, it is simpler to use a partial state observer when moving on to the next step of designing an observer for the system.

## 5.3 Partial State Observer Design

It is useful to analyze a partial state observer, assuming once again that V (membrane potential) is the measured variable. Assuming that V is measured, the dynamics for the observer and estimated variables become:

$$\dot{\hat{m}} = \overline{\lambda_m} \cosh\left(\frac{V - V_1}{2V_2}\right) \left(\frac{1}{2}\left(1 + \tanh\left(\frac{V - V_1}{V_2}\right)\right) - \hat{m}\right)$$

$$\dot{\hat{n}} = \overline{\lambda_n} \cosh\left(\frac{V - V_3}{2V_4}\right) \left(\frac{1}{2}\left(1 + \tanh\left(\frac{V - V_3}{V_4}\right)\right) - \hat{n}\right)$$

Then, the partial state observer Jacobian is simply:

$$\begin{bmatrix} -\overline{\lambda_m}\cosh(\frac{V - \hat{m}}{2V_2}) & 0 \\ 0 & -\overline{\lambda_n}\cosh(\frac{V - n}{2V_4}) \end{bmatrix}$$

Again, the range of the $\cosh(x)$ function for $-\infty < x < \infty$ is *always* positive.
This means that the eigenvalues of this diagonal matrix are always negative, because they are multiplied by $\overline{\lambda_m}$ and $\overline{\lambda_n}$, where both are positive constants, and then multiplied by -1. In order to affect the rate of convergence, on the other hand, it is necessary to feed back some kind of gain. Because only V is assumed to be measured, the easiest way is to feed back $k(\hat{V} - V)$.

## 5.4 Possible Approach to a Full-State Observer

Of course, this is only possible through the use of a full state observer, since V must be estimated along with m and n to use this form of feedback. This can be used to control the second and third lines of the equation in order to guarantee convergence, based on the bounds shown above for the off-diagonal terms. Namely, once specific feedback is used, contraction can be guaranteed for the observer through diagonal dominance, and the observer has all the correct pieces as solutions (e.g. $\hat{V}$ will converge to V).

$$C\dot{\hat{V}} = I - g_L V_L - g_{Ca}\hat{m}(\hat{V} - V_{Ca}) - g_K \hat{n}(\hat{V} - V_K) - k_1(\hat{V} - V)$$

$$\dot{\hat{m}} = \overline{\lambda_m} \cosh\left(\frac{\hat{V} - V_1}{2V_2}\right) \left(\frac{1}{2}\left(1 + \tanh\left(\frac{\hat{V} - V_1}{V_2}\right)\right) - \hat{m}\right) - k_2(\hat{V} - V)$$

$$\dot{\hat{n}} = \overline{\lambda_n} \cosh\left(\frac{\hat{V} - V_3}{2V_4}\right) \left(\frac{1}{2}\left(1 + \tanh\left(\frac{\hat{V} - V_3}{V_4}\right)\right) - \hat{n}\right) - k_3(\hat{V} - V)$$

The next step is to vary the gains $k_{1-3}$ in order to try to change the eigenvalues of the system in order to control the rate of convergence.

# 6. Hindmarsh-Rose Model[3]

## 6.1 Biological Motivation

In 1983, J.L. Hindmarsh and R. M. Rose developed a model composed of three coupled differential equations to describe neural activity. The work was motivated by the discovery of a cell in the pond snail *Lymnaea* which generated a burst after being depolarized by a short current pulse. This model describes the phenomenon of bursting in two-dimensional space. However, the model also incorporates a third dimension, which is a slow current that hyperpolarizes the cell. The variable x here refers to the membrane potential of the cell, while y and z represent the transport of ions across the cell membrane. The variable y represents the fast transport of sodium and potassium, while z represents the slower transport of other ions and is correlated to the bursting phenomenon.

6.2 MATHEMATICAL ANALYSIS

The Hindmarsh-Rose model is described by the following set of differential equations with three state variables:

$$\frac{dx}{dt} = y + bx^2 - ax^3 - z + I$$
$$\frac{dy}{dt} = c - dx^2 - y$$
$$\frac{dz}{dt} = r[s(x - x_R) - z]$$

The constant parameters are chosen to be at s=4, a=1, b=3, c=1, d=5, and r=10$^{-3}$.
This simplifies the Jacobian to the following form:

$$\begin{bmatrix} -3x^2 + 6x & 1 & -1 \\ -10 & -1 & 0 \\ 1/25 & 0 & -1/100 \end{bmatrix}$$

After trying various methods (e.g. diagonal dominance, negative definiteness, anti-symmetric negative diagonal) to assess the contraction properties of this system, the best (and simplest) method was determined to be the anti-symmetric with negative diagonal. Start with θ as the identity matrix. By varying the diagonal terms in θ, it is possible to affect the off-diagonal terms in the generalized Jacobian, $\dot{\theta}\theta^{-1} + \theta J \theta^{-1}$.

By testing out various values of the transformation matrix, and seeing how they affect the generalized Jacobian, the transformation matrix can then be chosen appropriately.
For example, from simple matrix algebra:

$$\text{If } \theta = \begin{bmatrix} c & 0 & 0 \\ 0 & 1 & 0 \\ 0 & 0 & 1 \end{bmatrix}, \text{then } \dot{\theta}\theta^{-1} + \theta J \theta^{-1}$$

$$= \begin{bmatrix} -3x^2 + 6x & 1 \cdot c & -1 \cdot c \\ -10/c & -1 & 0 \\ 1/25c & 0 & -1/100 \end{bmatrix}$$

$$\text{If } \theta = \begin{bmatrix} 1 & 0 & 0 \\ 0 & c & 0 \\ 0 & 0 & 1 \end{bmatrix}, \text{then } \dot{\theta}\theta^{-1} + \theta J \theta^{-1}$$

$$= \begin{bmatrix} -3x^2 + 6x & 1/c & -1 \\ -10 \cdot c & -1 & 0 \\ 1/25 & 0 & -1/100 \end{bmatrix}$$

$$\text{If } \theta = \begin{bmatrix} 1 & 0 & 0 \\ 0 & 1 & 0 \\ 0 & 0 & c \end{bmatrix}, \text{then } \dot{\theta}\theta^{-1} + \theta J \theta^{-1}$$

$$= \begin{bmatrix} -3x^2 + 6x & 1 & -1/c \\ -10 & -1 & 0 \\ c/25 & 0 & -1/100 \end{bmatrix}$$

By using a linear combination of the second two choices of transformation matrix, the simple choice of $\theta$ is:

$$\theta = \begin{bmatrix} 1 & 0 & 0 \\ 0 & 1/\sqrt{10} & 0 \\ 0 & 0 & 5 \end{bmatrix}$$

This yields a generalized Jacobian of:

$$\dot{\theta}\theta^{-1} + \theta J \theta^{-1} = \begin{bmatrix} -3x^2 + 6x & \sqrt{10} & -1/5 \\ -\sqrt{10} & -1 & 0 \\ 1/5 & 0 & -1/100 \end{bmatrix}$$

This matrix is anti-symmetric, meaning that $M = -M^T$. Now, all that remains is to ensure that the diagonals are negative for all values of x, y, and z. It is obvious that the significant choice is the variable (1,1) term, which depends only on x. The domain of concern is x=[0,2], since in this range of x values, the (1,1) term of the generalized Jacobian is positive.

To ensure that it is always negative, choose appropriate k gain values, namely, $k < -3$.

$$k = \begin{bmatrix} -10 & 0 & 0 \\ 0 & 0 & 0 \\ 0 & 0 & 0 \end{bmatrix}$$

Here, the gain k is arbitrarily chosen as -10, and as long as it is more negative than -3, can be varied depending on control power available to set this gain in the system. Thus, for all x, y, and z, this system as defined above is contracting.

6.3 FULL STATE OBSERVER DESIGN AND ANALYSIS

It is now possible to create an observer for this system. Assume that only the membrane potential (the x variable) can be measured, and it is necessary to reconstruct the state through the use of this measured variable.

Recall the form:
$$\dot{x} = f(x,t)$$
$$y' = h(x,t)$$

Here, $y'$ refers to the measured variable x. For this particular system, only x the membrane potential is assumed to be measured, while y and z are recreated through the use of a full-state observer.

$$\frac{d\hat{x}}{dt} = \hat{y} + \emptyset(\hat{x}) - \hat{z} - k_1(\hat{x} - x)$$
$$\frac{d\hat{y}}{dt} = \psi(\hat{x}) - \hat{y}$$
$$\frac{d\hat{z}}{dt} = r[s(\hat{x} - x_R) - \hat{z}]$$

where:
$$\emptyset(\hat{x}) = a\hat{x}^2 - \hat{x}^3$$
$$\psi(\hat{x}) = 1 - b\hat{x}^2$$

Then, the generalized Jacobian (of the form $\frac{d\dot{\hat{x}}}{d\hat{x}}$) incorporating the observer and using the aforementioned gain for $k_1$, becomes

$$\begin{bmatrix} -3\hat{x}^2 + 6\hat{x} - 10 & \sqrt{10} & -1/5 \\ -\sqrt{10} & -1 & 0 \\ 1/5 & 0 & -1/100 \end{bmatrix}$$

Thus, for all x, y, and z, this observer as defined above is contracting.

Because this is a full state observer, feedback using the estimate error for x can be used for all three lines of the observer, yielding the result of having control over the *rate* of convergence for the system, not just whether it converges.

6.4 SIMULATION
A simulation of the Hindmarsh Rose model in the following form yields the graph in Figure 2. The terms used are I=4, $x_r$=0, a=1, b=3, c=1, d=5, r=0.001, and s=4. The form of the model used is reproduced here for convenience. The difference between the model above and the one here is that I is not 0 (current applied). It does not affect the Jacobian and analyses above since it is a constant.

$$\dot{x} = \begin{bmatrix} y + bx^2 - ax^3 - z + I \\ c - dx^2 - y \\ r(s(x - x_r) - z) \end{bmatrix}$$

Figure 2a depicts the neuron bursting phenomenon derived from the Hindmarsh-Rose model definition and its simulation. Figure 2b plots two state variables against one another.

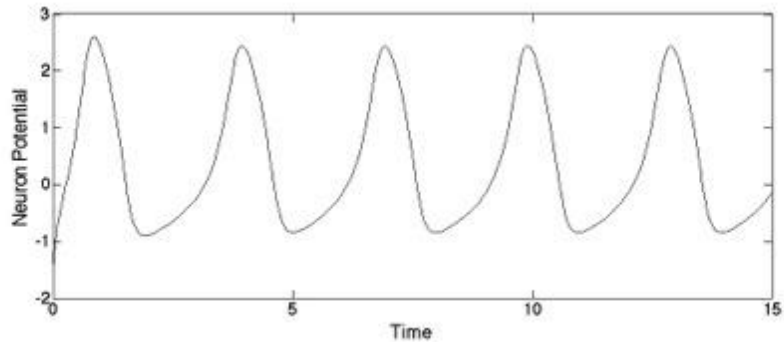

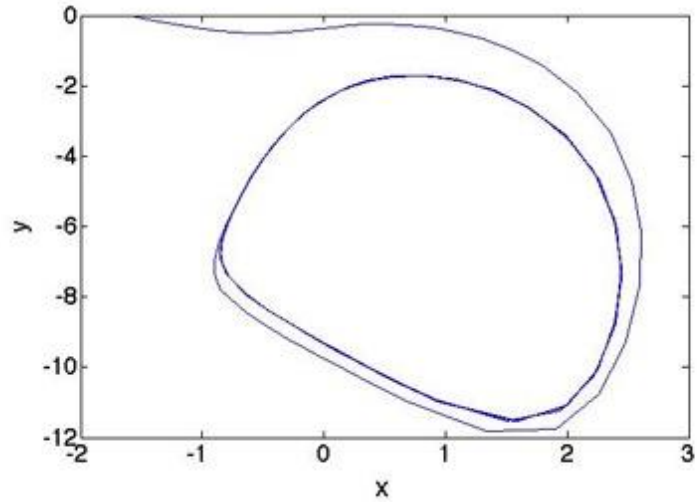

**Figure 2a**: MATLAB simulation of Hindmarsh-Rose model. **Figure 2b**: MATLAB plot of x and y variables against each other.

Figure 2b displays a limit cycle in the behavior of the system. Compare this simulation's results to a similar set of plots from the original Hindmarsh-Rose paper, to see the parallels, shown in Figure 3a and 3b.

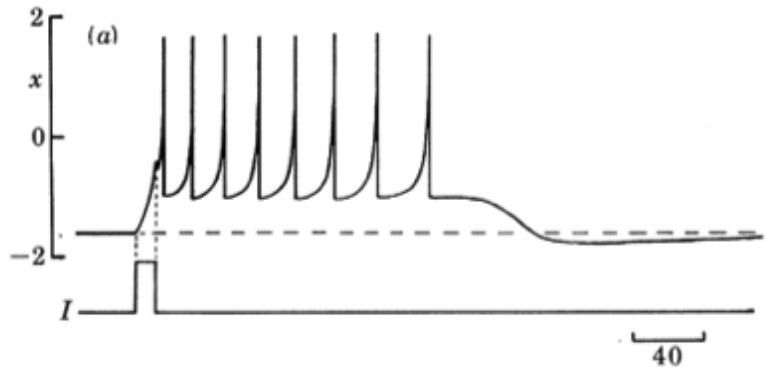

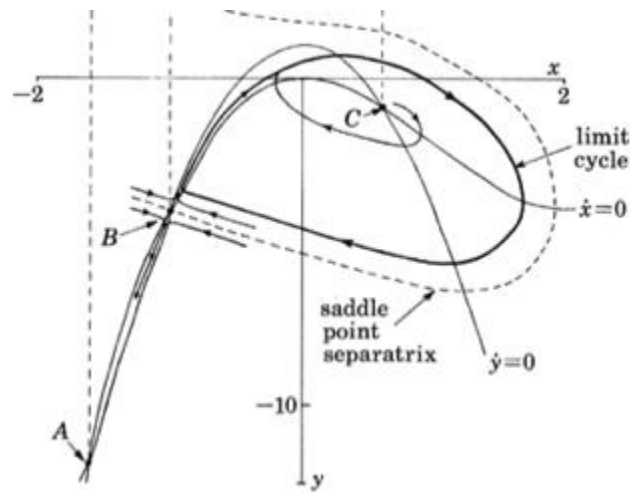

**Figure 3a**: Figure from original paper showing neuron spikes[3]. **Figure 3b**: Figure from original paper showing x and y coordinate plotted against one another.

## 7. Conclusions and Further Work

In conclusion, the four models analyzed using controls systems perspective provide some insights into the function of the various components in neuronal networks and their roles in signaling. The four-state Hodgkin-Huxley model can be simplified to smaller models such as the two-state Fitzhugh-Nagumo model to be studied in greater detail. Further work should address the possibility of a full-state observer for the Hodgkin-Huxley model, and also provide a better analysis of rate control for the Fitzhugh-Nagumo model. In addition, the possibility of a full-state observer for the Morris-Lecar model is open as well. This work provides a preliminary analysis and description of these four neural oscillator models from a nonlinear control systems approach, which can be expanded further to include improved detail, accuracy, and variety of methods and analysis techniques.

## 8. Acknowledgements and References

Many thanks to Professor Jean-Jacques Slotine at MIT, the advisor for this entire project who provided invaluable instruction and help to learn this material on nonlinear controls and neurobiology.